\documentclass[12pt]{article}
\usepackage{perpage} 
\MakePerPage{footnote}
\usepackage{amssymb}
\usepackage{amsmath,amsthm}
\usepackage{amsfonts}
\usepackage{graphicx}
\usepackage{braket}
\usepackage{graphics}
\usepackage{indentfirst}
\usepackage[colorlinks=true
  ,urlcolor=blue
  ,anchorcolor=blue
  ,citecolor=cyan
  ,filecolor= magenta
  ,linkcolor= magenta
  ,menucolor= magenta
  ,linktocpage=true
  ,pdfproducer=medialab
  ,pdfa=true
]{hyperref}

\begin{document}
 \begin{center}
        {\huge \textbf{Relativistic Magnetohydrodynamic Wave Excitation by 
        Laser Pulse in a Magnetized Plasma }}\\
        \end{center}
\begin{center}\textsc{\large
Zohreh Hashempour$^{a}$, 
Mehdi Nasri Nasrabadi\footnote{\href{mailto:mnnasrabadi@ast.ui.ac.ir }
{\texttt{mnnasrabadi@ast.ui.ac.ir} }
\textit{(Corresponding Author)}}$^{,a, b}$,
Nora Nassiri-Mofakham\footnote{\href{mailto:nnasiri@aeoi.org.ir}
{\texttt{nnasiri@aeoi.org.ir}}}$^{,c}$
} \&
\textsc{\large Hamidreza Daniali\footnote{\href{mailto:hrdl@aut.ac.ir}
{\texttt{hrdl@aut.ac.ir}}
}$^{,d}$.}
\centerline{$^a$ \it Department of Physics, University of Isfahan, 
81746-73441, Isfahan, Iran} 
\centerline{$^b$ \it Laboratory of Radiation Biology, 
Joint Institute for Nuclear 
Research, 141980 Dubna, Russia} 
   \centerline{$^c$ \it Fuel Cycle Research School, NSTRI, Tehran, Iran.}
\smallskip \centerline{$^d$ \it {Department of Physics, 
Amirkabir University of Technology, Tehran, Iran}}

\end{center}
\medskip
\begin{abstract}
\noindent
In the study of plasma, particularly in applications involving 
strong laser-plasma interactions, the propagation of a strong 
electromagnetic wave induces relativistic velocities in the electron flow. 
Given such conditions, the wave propagating through the 
plasma experiences modulational instability. 
In this paper, we investigate this instability using 
magnetohydrodynamic (MHD) equations. In the relativistic limit, 
the motion of ions can be neglected due to their significant inertia, 
allowing us to treat the ions as a background fluid. 
This simplification enables us to apply perturbation techniques 
to the electron fluid equations, leading to the derivation of the 
nonlinear wave equation in the form of the 
Nonlinear Schr\"odinger Equation (NLSE).
 We also explore the relationship 
between wave dispersion and the conditions for instability. 
We derive the maximum growth rate of the modulational instability 
and analyze its dependence on plasma parameters and wave intensity 
in the context of relativistic magnetized plasma, 
providing quantitative insights into the instability dynamics. 
Finally, we examine aspects of the perturbed NLSE using the 
Bogoliubov-Mitropolsky perturbation approach, 
treating real and imaginary coefficients separately, 
 which explicitly incorporates both Nonlinear Landau Damping 
(NLLD) and growth-damping effects.
\end{abstract}
\begin{flushleft}
\end{flushleft}
\thispagestyle{empty}
\clearpage
\section{Introduction}

Strong electromagnetic fields cause relativistic mass variation in electrons, 
which in turn induces modulational instabilities in wave propagation \cite{1}. 
This phenomenon is specific to electrons because the instability 
develops so quickly that ions, due to their greater inertia, 
cannot respond in time \cite{2}. Understanding the difference between 
electron and ion dynamics is crucial for studying the 
evolution of these instabilities in plasma environments \cite{3,4}.

Relativistic plasmas are found in various high-energy settings, 
such as fusion plasmas, astrophysical plasmas \cite{5,6}, 
the boundary layers of planetary magnetospheres \cite{7}, 
the Earth's radiation belts \cite{8}, and laboratory experiments 
involving laser-plasma interactions \cite{9,10}. 
The generation and amplification of magnetic fields in these 
contexts have long been of interest, with important applications 
in both plasma physics and astrophysics \cite{11}. 
Recent advancements in laser physics, particularly the 
development of Chirped Pulse Amplification (CPA) \cite{12}, 
have further accelerated research in these areas. 
CPA technology allows for the creation of ultra-intense, 
ultra-short laser pulses, which have significant implications for 
both scientific research and medical applications, 
such as surgery and diagnostics \cite{11}.

In plasma physics, CPA has opened new avenues for studying high-intensity 
laser-plasma interactions, with applications including the acceleration 
of high-energy ions, electrons, and positrons, as well as the generation 
of photon sources for fundamental physics 
experiments and radiotherapy \cite{13,14}. 
Investigating instabilities driven by laser-plasma 
interactions provides crucial insights into light emission 
in media where the refractive index deviates significantly 
from unity \cite{15,16}, which is essential for understanding 
astrophysical phenomena and cosmic ray acceleration. 
Recently, parametric instabilities induced by radiation 
forces in relativistic plasma dynamics have become a 
focal point of research, offering significant implications for 
understanding super-relativistic regimes \cite{17,18}.

The present work builds on recent advancements in generating laser 
pulses that can induce nonlinear effects, such as self-focusing, 
self-modulation, and parametric instabilities (see \cite{19} 
and references therein). These effects are observed in various 
physical processes, including inertial confinement fusion, 
higher-order harmonic generation, X-ray source development, 
and particle acceleration \cite{20,22}. 
Kumar et al. \cite{23} have shown that relativistic proton beams 
can stimulate large plasma resonances and accelerate electrons, 
underscoring the broader significance of this research.

In this work, we employ magnetohydrodynamic (MHD) equations \cite{24} 
to investigate instabilities arising from relativistic electron effects 
during the interaction of magnetized plasma with strong electromagnetic waves. 
By applying perturbation methods to the MHD equations describing 
ion motion, we derive the NLSE, 
which governs the interaction between plasma waves and electromagnetic fields. 
Our analysis establishes a maximum growth rate for 
modulational instabilities and examines the nonlinear 
behavior of wave amplitudes, 
considering both amplification and damping effects. 
Additionally, we derive electron-driven modes that exhibit characteristics 
analogous to Alfv\'en and magnetoacoustic waves \cite{25} under specific conditions, 
as a special case of our NLSE framework. 
Furthermore, we use the Bogoliubov-Mitropolsky perturbation approach to 
expand the NLSE solution, treating real and imaginary coefficients separately. 
This leads to NLLD for real coefficients 
and provides an alternative perspective on the 
growth-damping phenomenon for imaginary coefficients.

This work is organized as follows. 
In Sec. \ref{200}, we provide a concise review of the MHD equations and 
subsequently derive the NLSE pertinent to our configuration. 
Sec. \ref{300} is dedicated to a comprehensive analysis of modulational instability, 
including its dependence on plasma parameters and wave intensity, 
as well as its growth rates. 
In Sec. \ref{310}, we incorporate the nonlinear dynamics of wave 
amplitude to derive electron-driven modes with properties 
analogous to Alfv\'en and magnetoacoustic waves from the given configuration. 
Sec. \ref{400} focuses on the expansion of the NLSE, 
explicitly incorporating both NLLD and growth-damping effects, 
with the associated physical results detailed in 
Secs. \ref{410} and \ref{420}. Sec. \ref{500} devoted the conclusions.
\section{Physical model} \label{200}
Consider a high-density plasma through which a strong, 
linearly polarized electromagnetic wave with 
frequency $\omega_0$ and wave number $k_0$ is propagating. 
Due to the rapid relativistic modulation process occurring 
in the presence of intense electromagnetic radiation \cite{2}, 
ions can be treated as a background fluid. 
In this context, the equations of motion for the electron fluid are given by
\begin{eqnarray}
	\frac{\partial n}{\partial t} + \mathbf \nabla. 
	(n \mathbf V) &=& 0, \nonumber\\
	\frac{\partial\mathbf  P}{\partial t} + (\mathbf V . 
	\ \mathbf  \nabla) \mathbf P &=& - e \left( \mathbf E 
	+ \frac{1}{c} \mathbf V \times \mathbf  B\right),  \nonumber\\
	\mathbf \nabla \times \mathbf E &=& -\frac{1}{c} 
	\frac{\partial\mathbf B}{\partial t},  \nonumber\\
	\mathbf \nabla \times \mathbf B &=& \frac{4\pi}{c} \mathbf J 
	+ \frac{1}{c} \frac{\partial\mathbf E}{\partial t},  \nonumber \\
	\mathbf \nabla . \ \mathbf E &=& -4\pi e (n - n_0),  \nonumber\\
	\mathbf \nabla . \ \mathbf B &=& 0,
\end{eqnarray}
where $\mathbf J = ne\mathbf V$, and $c$ is the speed of light in vacuum and
\begin{eqnarray}
	\mathbf P = \frac{m \mathbf V}{\sqrt{1-\frac{V^2}{c^2}}} 
	= \gamma m \mathbf V. \nonumber
\end{eqnarray} 

With proper gauging, one can always express fields in terms of vector potential, 
$\mathbf A$,  and a scalar, $\phi$ as $\mathbf B = \mathbf \nabla \times \mathbf A$ 
and 
$\mathbf E = - \mathbf \nabla \phi - \frac{1}{c} \frac{\partial\mathbf A}{ \partial t}$. 
Therefore one obtains
\begin{eqnarray}\label{a}
	\frac{4\pi}{c} ne \mathbf V = \nabla^2 \mathbf A - \frac{1}{c^2} 
	\frac{\partial^2 \mathbf A}{\partial t^2}.
\end{eqnarray}
Let us now consider the vector potential as a wave with linear polarization, described by
\begin{eqnarray}
	\mathbf A = A(z,t) \cos \theta \hat e_x; \qquad\qquad \theta = k_0 z - \omega_0 t.
\end{eqnarray}
Utilizing $c \ \mathbf P = e \mathbf{A}$, we find
\begin{eqnarray}\label{b}
	\frac{ \omega_{p_0}^2}{2\omega_0 \gamma_0} A \frac{\delta n}{n_0} 
	= \frac{c^2}{2\omega_0} \frac{\partial^2 A}{\partial z^2} 
	+ i \left\{\frac{k_0 c^2}{\omega_0} \frac{\partial A}{\partial z} 
	+ \frac{\partial A}{\partial t} \right\}, \label{A}
\end{eqnarray}
where we have used $\gamma_0 = \sqrt{1+\frac{p^2}{mc^2}}$ and 
$n = n_0 + \delta n$ with density perturbation $\delta n$. 
The linear wave dispersion relation is also defined as 
$\omega_0^2 = k_0^2 c^2 + \frac{\omega_{P_0}^2}{\gamma_0}$ 
such $\omega_{P_0}^2 = \frac{4\pi n_0e^2}{m}$. 

To calculate the density perturbation $\delta n$, 
we consider the dynamics of plasma motion along 
the direction of wave propagation
\begin{eqnarray}
	\frac{\partial}{\partial t} \delta n + n_0 
	\frac{\partial}{\partial z} \delta V &=& 0, \label{A1}\\
	\frac{\partial}{\partial t} \delta V &=& 
	- \frac{e}{\gamma m} \left[\delta E - \frac{1}{c} v_0 \delta B \right], \label{A2}\\
	\frac{\partial}{\partial z}  \delta E &=& -4\pi e \delta n,  \label{A3}
\end{eqnarray}
where $v_0 = - \frac{e E_0 e^{i\theta}}{i m \omega_0}$. 
Also, $\delta B = \frac{  k_0 c E_0 e^{i\theta}}{\omega_0} $ 
is the magnetic field fluctuations. Applying the Fourier transformation to 
Eqs. \eqref{A1}-\eqref{A3} and assuming the relativistic factor $\gamma$,  
one obtains
\begin{eqnarray}
-i \omega_0 \delta n + i n_0 k_0 \delta V &=& 0, \label{fu-1} \\
-i \omega_0 \delta V &=& -\frac{e}{\gamma m} \left( \delta E 
- \frac{v_0}{c} \delta B \right), \label{fu-2} \\
i k_0 \delta E &=& -4\pi e \delta n. \label{fu-3}
\end{eqnarray}
Substituting the explicit form of unperturbed velocity and magnetic field fluctuations, 
together with Eq. \eqref{fu-3}, into Eq. \eqref{fu-2} yields
\begin{eqnarray}
	\delta V=\frac{4 \pi e^2}{\gamma m k_0 \omega_0} \delta n+\frac{k_0 e^2 
	E_0{ }^2 e^{2 i \theta}}{\gamma m^2 \omega_0{ }^3}. \label{deltanu}
\end{eqnarray}
Inserting Eq. \eqref{deltanu} into Eq. \eqref{fu-1} then leads to
\begin{eqnarray}
	\frac{\delta n}{n_0}=-\frac{k_0{ }^2 e^2 E_0{ }^2 
	e^{2 i \theta}}{m \omega_0{ }^2\left(n_0 4 \pi e^2
	-\gamma m \omega_0{ }^2\right)}.
\end{eqnarray}
Finally, using the definitions provided after Eq. \eqref{A}, we obtain
\begin{eqnarray}
	\frac{\delta n}{n_0} &=& \frac{e^2 (\gamma_0 \omega_0^2 
	- \omega^2_{P_0})}{m^2 c^2 \gamma_0 \omega_0^2 (\gamma \omega_0^2 
	- \omega^2_{P_0})} |E_0|^2 \cos(2\theta), \nonumber\\ \label{c}
	&=& -\frac{e^2 (\gamma_0 \omega_0^2 - \omega^2_{P_0})}{m^2 c^4 \gamma_0 
	(\gamma \omega_0^2 - \omega^2_{P_0})} |A_0|^2 \cos(2\theta),
\end{eqnarray}
which relates the density perturbation to the electromagnetic field amplitude. 
By substituting this into Eq. \eqref{b}, we obtain
\begin{eqnarray}\label{d}
	\frac{c^2}{2\omega_0} \frac{\partial^2 A}{\partial z^2} 
	+ i \left\{\frac{k_0 c^2}{\omega_0} \frac{\partial A}{\partial z} 
	+ \frac{\partial A}{\partial t} \right\} 
	= - \frac{\omega_P^2}{2\omega_0} \mathcal Q(\omega_0, \omega_{P_0}) |A|^2 A, 
	\end{eqnarray}
where we have defined
\begin{eqnarray}
	Q(\omega_0, \omega_{P_0}) = \frac{e^2 k_0^2}{m^2 c^2 \gamma_0 
	( \gamma \omega_0^2 - \omega_{P_0}^2)} \cos (2\theta).
\end{eqnarray}

By changing the variable $\xi \equiv z - \frac{k_0 c^2}{\omega_0}t$, 
$\tau \equiv t$ and working with the frame with the group velocity 
$\frac{k_0 c^2}{\omega_0}$, Eq. \eqref{d} takes the feature
\begin{eqnarray}\label{sch}
	i \frac{\partial A }{\partial \tau }+ \frac{c^2}{2\omega_0} 
	\frac{\partial^2 A}{\partial\xi^2} + \frac{\omega_{P_0}^2}{2\omega_0} 
	Q(\omega_0, \omega_{P_0}) |A|^2
 A = 0
 \end{eqnarray}
which is the conventional form of NLSE. 
It describes the propagation of relativistic modulational 
instability due to the excitation of hydrodynamic 
waves in the strong wave limit.

\section{Stability analysis}
\label{300}
In order to conduct a stability analysis, we examine the wave packet with 
an initial amplitude of $A_0$. Consequently, Eq. \eqref{sch} can be rewritten as 
\begin{eqnarray}\label{st}
	i \frac{\partial A}{\partial \tau} + \frac{c^2}{2\omega_0} 
	\frac{\partial^2 A}{\partial\xi^2} + \frac{\omega_{P_0}^2Q}{2 \omega_0}
	\bigg(|A|^2 - |A_0|^2\bigg) A = 0.
\end{eqnarray}
By considering, general solution, $A = \sqrt{g(\xi, \tau)} \ e^{i\sigma(\xi, \tau)}$ 
we get
\begin{eqnarray}
	&&\frac{\partial g}{\partial \tau} + \frac{c^2}{\omega_0} 
	\frac{\partial}{\partial\xi}\left(g \frac{\partial\sigma}{\partial\xi}\right) 
	= 0,  \nonumber\\ 
&&-\frac{\partial\sigma}{\partial \tau} + \frac{c^2}{4\omega_0 g_0}
\frac{\partial^2 g}{\partial\xi^2} -\frac{c^2}{8\omega_0 g^2} 
\left\{\left(\frac{\partial g}{\partial \xi} \right)^2 + 4g^2 
\left(\frac{\partial \sigma}{\partial \xi} \right)^2 \right\}
+ \frac{\omega_{P_0}^2 Q}{2\omega_0} 
(g- g_0) = 0.
\end{eqnarray}

For the sake of linearity, we apply a first-order perturbation to the given solutions, 
i.e., $g= g_0 + \varepsilon g_1(\xi, \tau) $ and 
$\sigma = \varepsilon \sigma_1 (\xi, \tau)$. 
Also, by assuming solutions of the form
$ g_1 , \sigma_1 \propto \exp (i K\xi - i \Omega \tau)$, one may obtain
\begin{eqnarray}
	-i \Omega g_1 - \frac{c^2 K^2}{\omega_0} g_0 \sigma_1 &=&0, \nonumber\\
	\left( \frac{\omega_{P_0}^2 Q}{2\omega_0} - \frac{c^2 K^2}{4\omega
	_0 g_0}\right) g_1 + i \Omega \sigma_1 &=& 0, 
\end{eqnarray}
which yields the following dispersion relation
\begin{eqnarray}\label{diss}
	\Omega^2 = \frac{c^2 K^2 g_0}{\omega_0} \left( \frac{c^2 K^2}{4\omega
	_0 g_0} -  \frac{\omega_{P_0}^2 Q}{2\omega_0} \right). 
\end{eqnarray} 
This equation indicates that the maximum growth rate of instability is
\begin{eqnarray}
	\Omega_{i_{\rm max}} = \frac{c^2 \omega_{P_0}^2 Q }{ \omega_0^3} |E_0|^2. 
\end{eqnarray}
Owing to  $\Omega= \Omega_r+ i\Omega_i$, the instability criterion is governed 
exclusively by the imaginary term. Precisely, when $\Omega_i<0$ ($\Omega_i>0$), 
the fluctuation wave attains stability (instability).

\subsection{Growth and damping effects}
\label{310}
Let us now consider the nonlinear dynamics of the wave amplitude, 
incorporating the effects of growth and damping \cite{26,27,28}. 
In our model, the assumption of static ions simplifies the MHD equations, 
focusing on relativistic electron dynamics. As we shall see, 
the stationary solutions resemble Alfv\'en and magnetoacoustic waves 
in their dispersion characteristics, though they are driven by 
electron motion rather than the collective ion-electron dynamics 
typical of classical MHD. 
This analogy arises due to the specific resonance condition 
$2k_1 = k_2 + k_3$ and the nonlinear interactions captured by the NLSE.
The wave excitation equation, Eq. \eqref{st}, can be rewritten as
\begin{eqnarray}\label{gd}
	i \left( \frac{\partial }{\partial \tau} + \tilde \gamma\right) A 
	+ \frac{c^2}{2\omega_0} \frac{\partial^2 A}{\partial\xi^2} 
	+ \frac{\omega_{P_0}^2Q}{2 \omega_0}\bigg(|A|^2 - |A_0|^2\bigg) A = 0, 
\end{eqnarray}
where $\tilde \gamma$ accounts for the growth and damping effects. 
By using the Fourier transform and the solution 
\begin{eqnarray}
	A(\xi , \tau) = \sum_{i=1}^3 A_i (\tau) e^{-i \phi_i (\xi , \tau)}
\end{eqnarray}
where
\begin{eqnarray}
	\phi_i (\xi , \tau)= k_i \xi - \omega_i \tau ; 
	\qquad\omega_i= -\frac{c^2}{2\omega_0} k_i^2,
\end{eqnarray}
with the resonance condition $2k_1 = k_2 + k_3$. After some calculations, we receive 
\begin{eqnarray}\label{20}
	\dot{A}_i\equiv \frac{d A}{d \tau} = - \tilde \gamma_i A_i 
	+ i\frac{\omega_{P_0}^2 Q}{2\omega_0}
	\left[ \left(\sum_{j\ne i}^3 |A_j|^2\right)  A_i + \Phi_i (A)\right],
\end{eqnarray}
where $\tilde \gamma_{2,3} = \tilde \gamma(k_{2,3})$,   
$\tilde \gamma_1 =- \tilde \gamma(k_1)$
 and the collective function $\Phi_i (A)$ possesses the feature
\begin{eqnarray}
	\Phi_i (A) = 2 A^\star_1 A_2 A_3 e^{2i\delta \tau} \delta_{i1} 
	+ (A_1^2 A^\star_3  \delta_{i2} + A_1^2 A^\star_2 
	 \delta_{i3}) e^{ - 2i\delta \tau}. 
\end{eqnarray}
in which $\delta \equiv 2^{-1} (\delta_2 + \delta_3)$ where 
$\delta_{2,3} = \omega_{2,3} - \omega_1$. Also, the double-index 
$\delta$ is the Kronecker delta function, i.e, $\delta_{ij} = 1$ 
where $i=j$ and zero otherwise.  

It should be noted that certain terms in $(|A|^2 - |A_0|^2) A$, 
such as $A_1 A_2 A^\star_3 e^{-i(\phi_1 + \phi_2 - \phi_3)}$, 
$A_1 A^\star_2 A_3 e^{-i(\phi_1 - \phi_2 + \phi_3)}$ and 
$A_2^2 A_3^\star e^{-i (2 \phi_2 - \phi_3)}$ are strongly 
suppressing due to the presence of the terms $2k_2 - k_1$ 
and $2k_2-k_3$, rendering these terms negligible.

Substituting the temporal part of the solution, 
$A_i (\tau) \equiv a_i (\tau) e^{i\psi_i (\tau)}$, 
into Eq. \eqref{20} and separating real and imaginary parts, we get
\begin{eqnarray}
	\dot a  &=& \tilde \gamma_1 a_1 + 2 a_1 a_2 a_3 
	\sin \theta,\nonumber\\
	\dot a_{2,3} &=& -\tilde \gamma_{2,3} a_{2,3} - a_1^2 a_{3,2} 
	\sin\theta,\nonumber\\
	\dot \theta &=& -2\delta + (a_3^2 + a_2^2 - 2 a_1^2) + \left[4a_2 a_3 
	- a_1^2\left(\frac{a_3}{a_2} + \frac{a_2}{a_3}\right)\right] \cos\theta,
\end{eqnarray}
in which $\theta(\tau) = 2\psi_1 - \psi_2 -\psi_3 -2\delta \tau$. 
Additionally, the stationary solution is achieved by setting 
$\dot a_1 = \dot A_2 = \dot \theta =0$
\begin{eqnarray}
	a_2^2 &=& - \frac{1}{2 \sin\theta_0} >0, \nonumber\\
	a_1^2 &=& - \frac{\tilde \gamma}{\sin\theta_0} >0, \nonumber\\
	\delta &=& (a_2^2 - a_1^2) + (2a_2^2-a_1^2) \cos \theta_0.  
\end{eqnarray}
Thus, we find
\begin{eqnarray}
	\sin \theta_0 = \frac{\delta (2 \tilde \gamma -1) \pm 2 (\tilde \gamma -1) 
	\sqrt{\delta^2 - \tilde\gamma + 3/4}}{2\bigg[\delta^2 
	+ (\tilde \gamma+1)^2 \bigg]} <0,
\end{eqnarray}
 and $\tilde \gamma>0$. 
Furthermore, to obtain a real solution while maintaining equilibrium, 
we get $\delta^2 > \tilde \gamma - \frac{3}{4}$. 
These relations correspond to electron-driven modes with properties analogous to 
Alfv\'en and magnetoacoustic waves \cite{25}.

\section{Perturbed NLSE} \label{400}
In this section, we extend the NLSE to include perturbations 
accounting for NLLD and growth-damping effects. 
These phenomena, modeled through real and imaginary coefficients respectively, 
capture distinct nonlinear dynamics in the relativistic electron fluid, 
providing a comprehensive framework for understanding wave 
stability and evolution in magnetized plasma.

The inverse scattering method has been employed to obtain the 
general solution of the NLSE in one spatial dimension. 
This solution consists of a superposition of stationary solutions, 
known as solitons, each associated with a discrete eigenvalue 
of the scattering potential, along with dispersive wave trains. 
Solitons can thus be viewed as nonlinear normal modes with exact 
solutions that can be expanded. Notably, solitons interact 
pairwise, leading to phase discontinuities. 
Considering the inertia of ions yields significant insights.

To study the effects of nonlinear phenomena, we introduce a perturbation 
term $\epsilon \Theta(A)$ to Eq. \eqref{sch} as
\begin{eqnarray} \label{psc}
	i \frac{\partial A }{\partial \tau }+ \frac{c^2}{2\omega_0} 
	\frac{\partial^2 A}{\partial\xi^2} + \frac{\omega_{P_0}^2}{2\omega_0} 
	Q(\omega_0, \omega_{P_0}) |A|^2
 A = \epsilon \Theta (A),
\end{eqnarray}
where $\epsilon \ll 1, c^2/2\omega_0$ and 
$\epsilon \ll \frac{\omega^2_{P_0}}{2\omega_0} Q(\omega_0, \omega_{P_0})$. 
Also we impose the condition $Q(\omega_0, \omega_{P_0})>0$. 
It should be noted that the term $ \Theta(A)$ represents additional 
nonlinear interactions in the 
relativistic electron fluid, which may arise from 
kinetic effects or phenomenological damping mechanisms. 
In Secs. \ref{410} and \ref{420}, 
we specify the form of $\Theta(A)$ for each case and provide 
physical interpretations of the resulting dynamics.
Now, by considering a perturbation around the unperturbed a soliton solution 
\begin{eqnarray}
	A_0 = 2G {\rm sech}\left\{ \frac{2 G\alpha}{mc^2}(\xi 
	- 2 W \tau) \exp\left[ \frac{2i}{c} \left( \frac{W \omega_0}{c} 
	(\xi - 2 W \tau) + t \left( W^2 \omega_0 
	+ \frac{(G\alpha)^2}{\omega_0 m^2}\right)\right)\right]\right\}
\end{eqnarray}
where we have defined $\alpha \equiv \omega_{P_0} e k \Psi(\theta)$ and 
$\Psi(\theta)\equiv \Big[\cos (2\theta)/ 2 \gamma_0 (\gamma \omega_0^2 
- \omega_{P_0}^2)\Big]^{1/2}$, with parameters $G$ and $W$ to be determined.
The perturbed solution can be expressed as 
$A = \mathcal R (\xi, \tau) \exp(i \Upsilon(\xi, \tau))$. Introducing 
\begin{eqnarray}
	\mathcal Z &\equiv& 2 G\left( \xi - 2\int_0^t W d\tau\right), \nonumber\\
	\Delta &\equiv& \frac{1}{\omega_0} \int_0^t  
	\left\{\left(\sqrt{2}\frac{ W \omega_0}{c}\right)^2 
	+ \left(\frac{G \alpha}{mc}\right)^2 \right\}d\tau,
\end{eqnarray}
and employing the Bogoliubov-Mitropolsky (BM) method \cite{29}, we expand
\begin{eqnarray}
	\mathcal R (\xi, \tau) &=& \mathcal R_0 (\mathcal Z, t_0, t_1) 
	+ \epsilon \mathcal R_1 (\mathcal Z, t_0) + \mathcal O (\epsilon^2),\nonumber\\
	\Upsilon (\xi, \tau) &=& \Upsilon_0 (\mathcal Z, t_0, t_1) 
	+ \epsilon\Upsilon_1 (\mathcal Z, t_0) +  \mathcal O (\epsilon^2),
\end{eqnarray}
in which, we have $t_0 = t$, $t_1= \epsilon t$ and one gets
\begin{eqnarray} \label{r_0}
	\mathcal R_0 (\mathcal Z, t_0, t_1) = 2 G {\rm sech} 
	\left(\frac{2G\alpha}{mc^2}\mathcal Z\right); \qquad\qquad \Upsilon_0 
	(\mathcal Z, t_0, t_1) = \frac{W \omega_0 }{G c^2}\mathcal Z + \Delta
\end{eqnarray} 

By substituting the perturbed solution to Eq. \eqref{psc}, 
we receive up to $\mathcal O (\epsilon^0)$ 
\begin{eqnarray}
	&&\frac{\partial \mathcal R_0}{\partial t_0} 
	- 4 WG \mathcal R_0 \frac{\partial \Upsilon_0}{\partial \mathcal Z} 
	+  \omega_0^{-1} (c G)^2\left[2 \frac{\partial\mathcal 
	R_0 \partial \Upsilon_0}{\partial \mathcal Z^2} + \mathcal R_0 
	\frac{\partial^2 \Upsilon_0}{\mathcal Z^2}\right] = 0, \\
	&&\mathcal R_0 \left(\frac{\partial}{\partial t_0} 
	+ 4 GW \frac{\partial}{\partial\mathcal Z}\right)\Upsilon_0 
	+ \frac{2 (cG)^2}{\omega_0}
	\left[\frac{\partial^2\mathcal R_0}{\partial\mathcal Z^2} 
	+ \mathcal R_0 
	\left(\frac{\partial\Upsilon_0}{\partial\mathcal Z}\right)^2\right] \nonumber 
	\\ &+& (2\omega_0 m^2 c^2)^{-1}\alpha^2 \frac{\partial^2 
	\Upsilon_0}{\partial \mathcal Z^2} = 0.
\end{eqnarray}
After a tedious computation, the following equation may be obtained in a 
unified form, up to order $\mathcal O (\epsilon)$ 
\begin{eqnarray}
	\left(\frac{\partial}{\partial t_0} - \mathcal A_{2\times 2}\right)
	\begin{bmatrix}
    \mathcal R_1   \\
    \Upsilon_1    
\end{bmatrix} = \begin{bmatrix}
     \mathcal S_1   \\
    \mathcal S_2   
\end{bmatrix},
\end{eqnarray}
with
\begin{eqnarray}
	\mathcal S_1 &=& -2 \frac{\partial\mathcal R_0}{\partial t_1} 
	+ \Im \Big[ \Theta(\mathcal R_0 , \Upsilon_0) e^{-i\Upsilon_0} \Big] ,\nonumber\\
	\mathcal S_2 &=& -2 \frac{\partial \Upsilon_0}{\partial t_1} 
	+ \Re \Big[ \Theta(\mathcal R_0 , \Upsilon_0) e^{-i\Upsilon_0} \Big].
\end{eqnarray}
Also, off-diagonal matrix $\mathcal A_{2\times 2}$ takes the feature
\begin{eqnarray}
	\mathcal A_{12} &=& \frac{4 c^2 G^3}{\omega_0}\left\{2 
	\frac{\partial}{\partial\mathcal Z}\left[{\rm sech} 
	\left(\frac{2G\alpha}{mc^2}\mathcal Z\right) 
	\frac{\partial}{\partial \mathcal Z}\right] 
	- {\rm sech} \left(\frac{2G\alpha}{mc^2}\mathcal Z\right) 
	\frac{\partial^2}{\partial \mathcal Z^2}\right\} ,\\
	\mathcal A_{21} &=& -2G \left( \frac{\alpha}{mc} \right)^2 
	\cosh \left(\frac{2G\alpha}{mc^2}\mathcal Z\right) 
	\Bigg\{ \left( \frac{\sqrt{2} mc^2}{\alpha} \right)^2 
	\frac{\partial^2}{\partial \mathcal Z^2} \nonumber\\
	&+& 6  {\rm sech} \left(\frac{2G\alpha}{mc^2}\mathcal Z\right) -1\Bigg\}.
\end{eqnarray} 

Given the presence of certain components in $\mathcal S \equiv \begin{bmatrix}
	\mathcal S_1   \\
    \mathcal S_2  
\end{bmatrix}$ contains components depending only on the slow time scale , $t_1$, 
it is inevitable that $\chi \equiv	\begin{bmatrix}
    \mathcal R_1   \\
    \Upsilon_1    
\end{bmatrix}$ will exhibit long-term behavior 
unless $\mathcal S$ is orthogonal to $\mathcal A \chi$, i.e.,
	\begin{eqnarray}\label{MM}
		\int_{-\infty}^{+ \infty} A(\mathcal Z)\  \mathcal S\  d\mathcal Z = 0,
	\end{eqnarray} 
if and only if $\int_{-\infty}^{+ \infty} A(\mathcal Z)\  \mathcal A
\chi d\mathcal Z = 0$.
Hence, we impose the condition $\mathcal A^* (A) = 0 $ 
where $\mathcal A^*$ is the adjoint operator of $\mathcal A$. 
To ensure that the solution of NLSE, $A = \begin{bmatrix}
    \lambda_A   \\
    \kappa_A    
\end{bmatrix}$, meets this condition, one can obtain
	\begin{eqnarray}\label{kap}
			&& \frac{\partial \lambda_A}{\partial\mathcal Z} 
			{\rm sech} \left(\frac{2G\alpha}{mc^2}\mathcal Z\right) 
			= \lambda_A \frac{\partial}{\partial \mathcal Z} {\rm sech} 
			\left(\frac{2G\alpha}{mc^2}\mathcal Z\right), \\
		&& -G c^2 \frac{\partial^2}{\partial \mathcal Z^2} 
		\Bigg[ \kappa_{A} \cosh \left(\frac{2G\alpha}{mc^2}\mathcal Z\right)	
		\Bigg]\nonumber\\ && \qquad + \frac{\kappa_{A} G}{2} \left(\frac{\alpha}{mc} 
		\right)^2 \cosh \left(\frac{2G\alpha}{mc^2}\mathcal Z\right)
		\bigg\{ 1- 6 {\rm sech} \left(\frac{2G\alpha}{mc^2}\mathcal Z\right) 
		\bigg\}=0.\label{lam}
	\end{eqnarray}
One can always utilize the above equation to determine the 
explicit forms of $\lambda_A$ and $\kappa_A$. 
However, using this notation and by performing dyadic product, 
Eq. \eqref{MM} acquires the feature 
\begin{eqnarray}
	\int_{-\infty}^{+ \infty} \lambda_A  \ \mathcal S_1\  d\mathcal Z 
	= \int_{-\infty}^{+ \infty} \kappa_A  \ \mathcal S_2\  d\mathcal Z = 0.
\end{eqnarray} 
Consequently, the independent parameters $G$ and $W$ 
exhibit a temporal variation on the time scale $t_1$
\begin{eqnarray}\label{GG}
	\frac{\partial G}{\partial t_1} &=& \frac{1}{4} \left(\frac{\alpha}{mc^2} \right)^2 
	\Im \int_{-\infty}^{+\infty} \Theta(\mathcal R_0 , \Upsilon_0) 
	e^{-i\Upsilon_0} {\rm sech} \left(\frac{2G\alpha}{mc^2}\mathcal Z\right) 
	d \mathcal Z,\\
\frac{\partial W}{\partial t_1} &=& -\left(\frac{\alpha}{2\sqrt{\omega_0}mc} 
\right)^2 \Re \int_{-\infty}^{+\infty} \dfrac{\tanh 
\left(\frac{2G\alpha}{mc^2}\mathcal Z\right)}{\cosh \left(\frac{2G\alpha}{mc^2}\mathcal 
Z\right)} \Theta(\mathcal R_0 , \Upsilon_0) e^{-i\Upsilon_0}  d \mathcal Z	.\label{WW}
\end{eqnarray}
 In order to express our solution in terms of a single time, we can 
 substitute $t_0 = t$ and $t_1= \epsilon t$.
\subsection{Phenomenological NLLD} \label{410}
NLLD is typically derived from kinetic theory, 
where it describes wave-particle interactions leading to energy transfer 
between waves and particles. 
In this fluid model, we introduce NLLD phenomenologically to capture 
similar damping effects in the relativistic electron fluid, 
following approaches such as those in Ref. \cite{29}. 
While a rigorous kinetic derivation is beyond the scope of this work, 
we approximate NLLD as an effective damping term in the NLSE, 
with the form given below, and leave the detailed justification for future studies.
The NLLD term, given by
\begin{eqnarray}
	\Theta (\mathcal R_0, \Upsilon_0) = - \frac{1}{\pi} 
	\mathcal P \int_{-\infty}^{+\infty} d \mathcal Z (\mathcal Z 
	- \mathcal Z') \mathcal R^2(\mathcal Z', t) R(\mathcal Z, t) 
	\exp(i \Upsilon (\mathcal Z, t)),
\end{eqnarray}
approximates the energy transfer due to wave-particle interactions in the 
fluid framework, leading to the conservation of quanta and 
soliton motion. By plugging it into Eqs. \eqref{GG} and \eqref{WW}, we find
\begin{eqnarray}\label{lan1}
	\frac{\partial W}{\partial t} &=& \frac{8 \epsilon G^3}{\omega_0 \pi} 
	\left(\frac{\alpha}{2mc}\right)^2 \nonumber\\ &\times& \mathcal P 
	\int_{-\infty}^{+\infty}
	\int_{-\infty}^{+\infty} 
	\frac{d \mathcal Z'\ d \mathcal Z }{\mathcal Z - \mathcal Z'} 
	\sinh \left(\frac{2G\alpha}{mc^2}\mathcal Z\right) 
	{\rm sech}^3 \left(\frac{2G\alpha}{mc^2}\mathcal Z\right){\rm sech}^2 
	\left(\frac{2G\alpha}{mc^2}\mathcal Z'\right), \\\label{lan2}
	&=& \epsilon \omega_0^{-1} \left( \frac{2 \alpha}{\pi mc}\right)^2 
	\Gamma(4) \zeta(3),\\
	\frac{\partial G}{\partial t} &=& 0. \label{lan3}
\end{eqnarray}
Eqs. \eqref{lan1}-\eqref{lan3} may also be derived by 
substituting the solution with time-dependent $G$ and $W$ into the equations 
for the time derivative of the number of quanta, momentum, and energy. 
NLLD is widely recognized for its unique feature of conserving the number of quanta, 
resulting in the conversion of a higher-frequency wave quanta into one with a 
lower frequency. 

Furthermore, according to Eqs. \eqref{lan1}-\eqref{lan3}, 
it is evident that a soliton that is initially stationary would initiate 
motion as a result of the NLLD. While the Bogoliubov-Mitropolsky method is a 
standard approach for solving nonlinear equations, its application here to the 
NLSE in the context of relativistic magnetized plasma, incorporating both NLLD 
and growth-damping effects, provides new insights into the 
nonlinear dynamics of wave propagation

\subsection{Growth and damping effects as a complex coefficient}
\label{420}
Let us return to Sec. \ref{310} to investigate the growth and damping effects 
as a perturbation to the NLSE. By comparing Eq. \eqref{gd} and 
Eq. \eqref{psc}, it becomes evident that the growth and damping effects
can be introduced as a perturbation to NLSE as
\begin{eqnarray}
	\Theta (A) = -i \tilde \gamma A + \frac{1}{2\omega_0} 
	\left(\frac{\alpha}{mc} \right)^2 |A_0| A. 
\end{eqnarray}

Substituting $\Theta (A)$ into Eq. \eqref{GG} and Eq. \eqref{WW} 
and evaluating integrals, one finds
\begin{eqnarray}\label{Gg}
	\frac{\partial G}{\partial t} &=& -2 G \epsilon \left[ \tilde \gamma 
	+ \frac{i}{3\omega_0} \left(\frac{2\alpha}{mc}\right)^2 G^2\right], \\
	\frac{\partial W}{\partial t} &=& 0.\label{Wd} 
\end{eqnarray}
Eq. \eqref{Gg} describes the interplay between linear growth 
and nonlinear stabilization of the soliton's amplitude. 
When considering collisional and Landau damping, 
the soliton's velocity remains unchanged by collisional 
damping if $\tilde \gamma > 0$. 
This contrasts with NLLD, where an initially stationary soliton accelerates. 
These results can be extended to scenarios involving the transition 
from soliton solutions to shock waves in lossless 
but slightly inhomogeneous plasmas.
\section{Conclusions}  \label{500}
In this study, we investigated the instability resulting from relativistic 
effects on electrons during the interaction of strong electromagnetic waves 
with plasma, using MHD equations. By applying perturbation methods to the MHD 
equations governing ion motion, we derived the NLSE, which captures the 
interaction between electromagnetic and plasma waves. 
We derived an analytical expression for the maximum growth rate of the 
modulational instability, providing quantitative insights into its 
dependence on plasma parameters and wave intensity in 
relativistic magnetized plasma.
In specific cases, our analysis yields electron-driven 
modes with properties analogous to Alfvén and magnetoacoustic waves, 
arising from the nonlinear dynamics of the relativistic electron fluid.

The general solution of the NLSE consists of a superposition of 
stationary solutions, or solitons, each corresponding to discrete 
eigenvalues of the scattering potential, along with dispersive wave trains. 
A key finding of this work is that solitons can be viewed as nonlinear normal modes. 
We expanded the NLSE solution using the Bogoliubov-Mitropolsky perturbation method, 
treating real and imaginary coefficients separately. 
This approach systematically addresses the real coefficient case, 
explicitly incorporating NLLD, while the imaginary coefficient case explicitly 
incorporates the growth-damping effect, providing insights into their 
interplay within the NLSE framework.
\section*{Acknowledgments}
The authors would like to thank Sasan Rezaee and Mohammad 
Nikoosefat for fruitful discussions, 
and Alireza Javadi, without whom this work could never be done.


\end{document}